\documentclass[11pt]{article}
\usepackage[pctex32]{graphics}

\def\ben{\begin{equation}}
\def\een{\end{equation}}

\def\nn{\nonumber} \def\bd{\begin{document}} \def\ed{\end{document}}
\def\ds{\documentstyle} \let\fr=\frac \let\bl=\bigl \let\br=\bigr
\let\Br=\Bigr \let\Bl=\Bigl
\let\bm=\bibitem
\let\na=\nabla
\let\pa=\partial \let\ov=\overline
\newcommand{\be}{\begin{equation}}
\newcommand{\ee}{\end{equation}}
\def\ba{\begin{array}}
\def\ea{\end{array}}
\def\ft#1#2{{\textstyle{\frac{\scriptstyle #1}{\scriptstyle #2} } }}
\def\fft#1#2{{\frac{#1}{#2}}}
\def\del{\partial}
\def\vp{\varphi}
\def\sst#1{{\scriptscriptstyle #1}}
\def\oneone{\rlap 1\mkern4mu{\rm l}}
\def\td{\tilde}
\def\wtd{\widetilde}
\def\ie{{\it i.e.\ }}
\def\dalemb#1#2{{\vbox{\hrule height .#2pt
        \hbox{\vrule width.#2pt height#1pt \kern#1pt
                \vrule width.#2pt}
        \hrule height.#2pt}}}
\def\square{\mathord{\dalemb{6.8}{7}\hbox{\hskip1pt}}}
\newcommand{\ho}[1]{$\, ^{#1}$}
\newcommand{\hoch}[1]{$\, ^{#1}$}
\newcommand{\bea}{\setlength\arraycolsep{2pt} \begin{eqnarray}}
\newcommand{\eea}{\end{eqnarray}}
\newcommand{\ra}{\rightarrow}
\newcommand{\lra}{\longrightarrow}
\newcommand{\Lra}{\Leftrightarrow}
\newcommand{\bp}{\tilde \beta^\prime}
\newcommand{\tr}{{\rm tr} }
\newcommand{\Tr}{{\rm Tr} }
\def\0{{\sst{(0)}}}
\def\1{{\sst{(1)}}}
\def\2{{\sst{(2)}}}
\def\3{{\sst{(3)}}}
\def\4{{\sst{(4)}}}
\def\5{{\sst{(5)}}}
\def\6{{\sst{(6)}}}
\def\7{{\sst{(7)}}}
\def\8{{\sst{(8)}}}
\def\m{{\sst{(m)}}}
\def\n{{\sst{(n)}}}
\def\cA{{{\cal A}}}
\def\cB{{{\cal B}}}
\def\cF{{{\cal F}}}
\def\cG{{{\cal G}}}
\def\cH{{{\cal H}}}
\def\tV{\widetilde V}
\def\tW{\widetilde W}
\def\tH{\widetilde H}
\def\tE{\widetilde E}
\def\tF{\widetilde F}
\def\tA{\widetilde A}
\def\im{{{\rm i}}}
\def\tY{{{\wtd Y}}}
\def\ep{{\epsilon}}
\def\vep{{\varepsilon}}
\def\bD{{{\bar D}}}
\def\R{{{\mathbb R}}}
\def\C{{{\mathbb C}}}
\def\H{{{\mathbb H}}}
\def\CP{{{\mathbb C}{\mathbb P}}}
\def\RP{{{\mathbb R}{\mathbb P}}}
\def\Z{{{\mathbb Z}}}
\def\bA{{{\mathbb A}}}
\def\bB{{{\mathbb B}}}
\def\bC{{{\mathbb C}}}
\def\bD{{{\mathbb D}}}
\def\bE{{{\mathbb E}}}
\def\bZ{{{\mathbb Z}}}
\def\Re{{{\frak{Re}}}}
\def\Im{{{\frak{Im}}}}
\def\cosec{{\,\hbox{cosec}\,}}
\def\Gm{{\Gamma_{\!\! -}}}
\def\Gp{{\Gamma_{\!\! +}}}
\def\stan{{standard }}
\def\nonstan{{supernumerary }}
\def\p{{\partial}}
\def\kdel#1{{\fft{\del}{\del#1}}}

\def\bog{{Bogomolny }}
\def\om{{\omega}}

\newcommand{\nnr}{\nonumber \\}
\newcommand{\pd}{\partial}
\newcommand{\ud}{\textrm{d}}
\newcommand{\dTH}{T^{\prime \, 0}_\textrm{H}}
\newcommand{\dOi}{\Omega^{\prime \, 0}_i}
\newcommand{\bx}{{\bf x}}

\begin{document}

\vspace{5mm}
\begin{center}
{\Large \bf Thermodynamics of Ho\v{r}ava-Lifshitz black holes}
\vspace{12mm}

{\large   Yun Soo Myung \footnote{e-mail
 address: ysmyung@inje.ac.kr} and Yong-Wan Kim \footnote{e-mail
 address: ywkim65@gmail.com}}
 \\
\vspace{10mm} {\em Institute of Basic Science and School of
Computer Aided Science \\ Inje University, Gimhae 621-749, Korea}
\end{center}

\begin{center}

\underline{Abstract}
\end{center}

  We study  black holes in the Ho\v{r}ava-Lifshitz gravity  with a parameter $\lambda$.
  For $1/3 \le \lambda < 3$, the black holes behave the
  Lifshitz black holes with dynamical exponent $0 < z \le 4$, while for $\lambda > 3$, the black holes behave the
  Reissner-Nordstr\"om  type black hole in asymptotically flat spacetimes.
  Hence, these all are quite different from the Schwarzschild-AdS black
  hole of Einstein gravity. The temperature, mass, entropy, and
  heat capacity are derived for investigating  thermodynamic
  properties of these black holes.

\vspace{15pt}

\thispagestyle{empty}





\newpage
\section{Introduction}
Ho\v{r}ava has proposed a renormalizable theory of gravity at a
Lifshitz point~\cite{ho1},  which  may be regarded as a UV
complete candidate for general relativity.  Recently, the
Ho\v{r}ava-Lifshitz gravity theory has been intensively
investigated in~\cite{ho2,ho3,VW,klu,Nik,Nas,Iza,Vol,CH} and its
cosmological applications have been  studied in
~\cite{cal,TS,muk,Bra,pia,gao}.

Introducing the ADM formalism where the metric is parameterized
\cite{adm}
\be ds_{ADM}^2= - N^2  dt^2 + g_{ij} \Big(dx^i - N^i dt\Big)
\Big(dx^j - N^j dt\Big)\,, \ee
the Einstein-Hilbert action can be expressed as
\be \label{Eins} S_{EH} = \fft{1}{16\pi G} \int d^4x \sqrt{g} N
\Big(K_{ij} K^{ij} - K^2 + R - 2\Lambda\Big)\,, \ee
where $G$ is Newton's constant and extrinsic curvature $K_{ij}$
takes the form
\be K_{ij} = \fft{1}{2N} \Big(\dot g_{ij} - \nabla_i N_j -
\nabla_j N_i\Big)\,. \ee
Here, a dot denotes a derivative with respect to $t$.

   The action of the $z=3$  Ho\v{r}ava-Lifshitz  with a parameter $\lambda$ is given by
\bea%
S_{HL}&=&\int dtd^3\bx\, \Big({\cal L}_0 + {\cal L}_1\Big)\,,\nn\\
{\cal L}_0 &=& \sqrt{g}N\left\{\frac{2}{\kappa^2}(K_{ij}K^{ij}
\label{action1}-\lambda K^2)+\frac{\kappa^2\mu^2(\Lambda_W R
  -3\Lambda_W^2)}{8(1-3\lambda)}\right\}\,,\\ {\cal L}_1&=&
\sqrt{g}N\left\{\frac{\kappa^2\mu^2
(1-4\lambda)}{32(1-3\lambda)}R^2 -\frac{\kappa^2}{2w^4}
\left(C_{ij} -\frac{\mu w^2}{2}R_{ij}\right) \left(C^{ij}
-\frac{\mu w^2}{2}R^{ij}\right)\right\}\,.\label{action2}
\eea%
where $C_{ij}$ is the Cotton tensor defined by
\be C^{ij}=\epsilon^{ik\ell}\nabla_k\left(R^j{}_\ell
-\frac14R\delta_\ell^j\right).\label{def.K.C} \ee
Comparing ${\cal L}_0$ with Eq.(\ref{Eins}) of general relativity,
the speed of light, Newton's constant,  the cosmological constant,
paramter $\lambda$  are  determined  by
\be c=\fft{\kappa^2\mu}{4}
\sqrt{\fft{\Lambda_W}{1-3\lambda}}\,,\qquad
G=\fft{\kappa^2}{32\pi\,c}\,,\qquad \Lambda=\ft32 \Lambda_W\,
,\qquad \lambda=1\,.\label{cg} \ee
The equations of motion were derived in \cite{LMP} and \cite{KK},
but we do not write  them  due to the length.

In this work, we investigate the Ho\v{r}ava-Lifshitz black hole
solutions and their thermodynamic properties.

\section{Lifshitz black holes}

Considering $N^2=\tilde{N}^2f(r)$ and $N^i=0$, a spherically
symmetric solution could  be obtained with a metric ansatz
proposed by L\"u-Mei-Pope (LMP)~\cite{LMP,CCO,CLS,CY}
\be \label{ssm} ds_{\rm LMP}^2 = - \tilde{N}^2(r) f(r)\,dt^2 +
\fft{dr^2}{f(r)} + r^2 (d\theta^2 +\sin^2\theta d\phi^2)\,, \ee
which implies that \be K_{ij}=0,~C_{ij}=0. \ee The first condition
of $K_{ij}=0$ means that the embedding is trivial for a
spherically symmetric, static  solution  and the non-zero Cotton
tensor $C_{ij}$ is not necessary to obtain  a spherically
symmetric solution. Taking the Lagrangian ${\cal L}_0$ only, we
obtain the Schwarzschild-AdS (SAdS) black hole  whose metric
function is given by
\be f=1 - \fft{\Lambda_W}{2} r^2 - \fft{m}{r}\,\label{adsbh} \ee
with $\tilde{N}^2=1$. The simplest way  to obtain a spherically
symmetric  solution for the Ho\v{r}ava-Lifshitz gravity   is to
substitute the metric ansatz (\ref{ssm}) into the action. Then,
let us  vary the reduced action with respect to  functions
$\tilde{N}$ and $f$. This is possible because the metric ansatz
shows all the allowed singlets which are compatible with the
$SO(3)$ action on the $S^2$. The reduced Lagrangian is given by
\be \label{react} {\cal L}= \fft{ \kappa^2\mu^2 \Lambda_W\tilde{N}
}{8(1-3\lambda)}\Bigg[2(1  -f - r f')- 3\Lambda_W r^2+
\fft{\lambda-1}{2\Lambda_W} f'^2 - \fft{2\lambda (f-1)}{\Lambda_W
r}f' + \fft{(2\lambda-1)(f-1)^2}{\Lambda_W r^2}\Bigg]. \ee
The first two terms come from ${\cal L}_0$, while the remaining
terms come from higher order Lagrangian ${\cal L}_1$ with
$C^{ij}=0$. Thus, in the limit of $-\Lambda_W \to \infty$, we
expect to find the SAdS black hole (\ref{adsbh}). However, this
limit is the other limiting case, compared to the asymptotically
flat spacetime limit of $-\Lambda_W \to 0$ in anti-de Sitter
spacetimes. Hence, we should be careful to treat the limiting case
of $-\Lambda_W \to \infty$ in Ho\v{r}ava-Lifshitz gravity.

 Introducing a newly radial coordinate
$x=\sqrt{-\Lambda_W}r$, we have LMP black hole solutions where $f$
and $\tilde{N}$ are determined to be \bea  \label{sol1} f&=&1 +
x^2-\,
\Big(\sqrt{-\Lambda_W} mx\Big)^{p_\pm(\lambda)},~p_\pm(\lambda)=\frac{2\lambda\pm\sqrt{6\lambda-2}}{\lambda-1},\\
\label{sol2}
\tilde{N}&=&x^{q_\pm(\lambda)},~q_\pm(\lambda)=-\frac{1+3\lambda\pm2\sqrt{6\lambda-2}}{\lambda-1}
\,,\label{nf} \eea where $m$ is a mass parameter related to the ADM
mass\footnote{Our definition of mass parameter is slightly different
from Cai-Cao-Ohta~\cite{CCO} where a mass parameter $m$ was defined
by $\sqrt{m} r^{p(\lambda)}$. In this case, the mass dimension seems
to be incorrect. Also, the previous definition of $mx^{p(\lambda)}$
has a problem to derive the entropy.}. For the solution to be real,
it requires $\lambda \ge 1/3$. We mention that $p_+$ and $q_+$
exponents are discarded because they are singular at $\lambda=1$. In
this work, thus we make replacements: $p_-(\lambda) \to p(\lambda) $
and $q_-(\lambda) \to q(\lambda)$ with $-\Lambda_W=1/l^2=1$ for
simplicity.  Then, $x=r/l$ becomes a convenient variable for
describing the LMP black holes. Hereafter we call the
Ho\v{r}ava-Lifshitz black holes as the LMP black holes.
\begin{figure}[t!]
   \centering
   \includegraphics{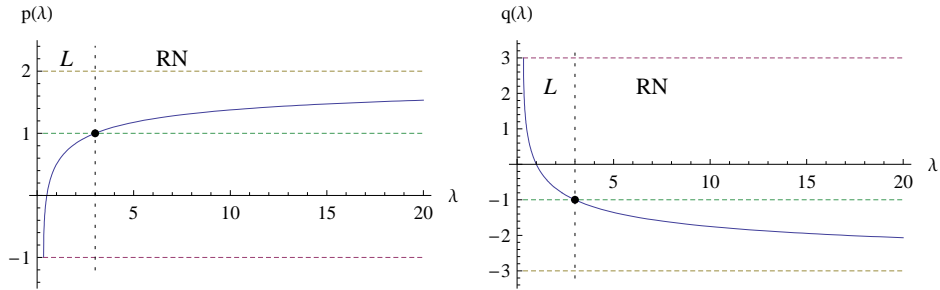}
\caption{Exponent graphs of  $p(\lambda)$ and $q(\lambda)$. Here
we observe $p(1/3)=-1,~p(1/2)=0,~p(1)=1/2,~p(3)=1(\bullet),$ and
$p(\infty)=2$, while we observe
$q(1/3)=3,~q(1/2)=1,~q(1)=0,~q(3)=-1(\bullet),$ and $
q(\infty)=-3$. For $1/3 \le \lambda < 3$ (L), the Lifshitz black
holes $(0 < z \le 4)$ appear, while the Reissner-Nordstr\"om type
black holes (RN) are shown for $\lambda >3$. } \label{fig.1}
\end{figure}

Before proceeding, since the solution of (\ref{sol1})-(\ref{sol2})
is similar to the charged dilaton solution in three dimensional
AdS spacetimes~\cite{CM,MKP}, we may  use this  idea to study and
Ho\v{r}ava-Lifshitz black holes and their thermodynamics.
 As is
shown in Fig. 1, for $1/3 \le \lambda < \infty$, we have  two
bounds of $-1\le p(\lambda)< 2$ and  $-3< q(\lambda) \le 3$. The
first bound implies that the $m$-term plays a role of the mass
term because its exponent is always less than ``2", the second
term of AdS spacetimes. Importantly, the latter bound appears
because higher order curvature terms like $R^2$ and $R_{ij}R^{ij}$
are present, reflecting the Ho\v{r}ava-Lifshitz gravity. We would
like to mention  the three cases of $\lambda=1/3$, $\lambda=1/2$,
and $\lambda=1$. In the case of $\lambda=1/3$, we have
$f=1+x^2-\frac{l^2}{ m r }$ and $\tilde{N}=x^3$ with
$x=\frac{r}{l}$,
 whereas  for SAdS case,  $f=1+x^2-\frac{m}{r}$ and
 $\tilde{N}=1$. Hence, we expect that its
 thermodynamic property is  different from that of SAdS even though the metric functions $f$ are the same.
 For $\lambda=1/2$, we have $f=x^2$ and $\tilde{N}=x$ which may
 imply that its thermodynamics is marginally defined.
 Also, the case of
$\lambda=1$ seems to be familiar  as $f=1+x^2-\frac{\sqrt{mr}}{l}$
and $~\tilde{N}=1$. However, the LMP black holes in (\ref{sol1})
are not yet  completely classified and understood.

In order to understand the LMP black holes, it is necessary to
introduce  Lifshitz black holes whose asymptotic form takes the form
with dynamical exponent $z$~\cite{KLM,DT,Mann,BBP1,BBP2,BM,AGGH} \be
\label{lifsh} ds^2_{\rm Lifshitz}=- x^{2z}
F(r)dt^2+\frac{1}{H(r)x^2}dr^2+r^2d\Omega^2_{d-2}, \ee where $F(r)$
and $H(r)$ are functions of  radial coordinate $r$ with \be \lim_{r
\to \infty}F(r)=\lim_{r \to \infty}H(r)=1. \ee Comparing the LMP
black holes (\ref{sol1}) with the Lifshitz black holes (\ref{lifsh})
shows  the correspondence \be \tilde{N}^2f=x^{2(q+1)}\frac{f}{x^2}
\to x^{2z}F(r),~~f \to H(r)x^2 \ee which leads to two relations \be
z=q+1,~~F(r)= H(r)=\frac{f}{x^2}. \ee

\begin{table}

 \caption{Summary for the Lifshitz black holes in the Ho\v{r}ava-Lifshitz black hoes.}
\begin{tabular}{|c|c|c|c|c|}
  \hline
  $\lambda$ & $\frac{1}{3}$ & $\frac{1}{2}$  & 1 & 3 \\
  \hline
  $q$ & 3 & 1 & 0 & $-1$ \\ \hline
  $z$ & 4 & 2 & 1 & 0 \\
  \hline
  $p$ & $-1$ & 0 & $\frac{1}{2}$ & 1 \\
   \hline
  $f(r)$ & $1+x^2-\frac{l^2}{ m r }$  & $1-m+x^2$ &$1+x^2-\frac{\sqrt{mr}}{l}$  &
  $1+x^2-\frac{mr}{l^2}$ \\
   \hline
  $N^2(r)=\tilde{N}^2f$ & $x^6(1+x^2-\frac{l^2}{ m r })$  & $x^2(1-m+x^2)$ &$1+x^2-\frac{\sqrt{mr}}{l}$  &
  $1-\frac{m}{r}+\frac{1}{-\Lambda_W r^2}$ \\
  \hline
\end{tabular}
\end{table}
 As is  shown in Table 1, we have four interesting cases with
Lifshitz asymptotics.  For $1/3 \le \lambda \le 1/2$, the
Ho\v{r}ava-Lifshitz (LMP) back holes are similar to the
non-rotating BTZ black holes, while for $1/2 \le \lambda < 3$,
they behave the rotating BTZ black hole. They all belong to the
Lifshitz black holes  with  $0< z \le 4$. However, we observe that
for $\lambda=3$, the lapse function $N^2$ takes the metric
function for the Reissner-Nordstr\"om black hole when interpreting
$\frac{1}{-\Lambda_W}$ to be the charge of $Q^2$. Hence, for
$\lambda
>3$, they seem to be completely different from the Lifshitz black
holes, Reissner-Nordstr\"om type black holes. In order to confirm
the mentioned conjecture, we need to study thermodynamic
properties of the LMP black holes.

\section{Thermodynamics of  Ho\v{r}ava-Lifshitz black
holes}
\begin{figure}[t!]
   \centering
   \includegraphics{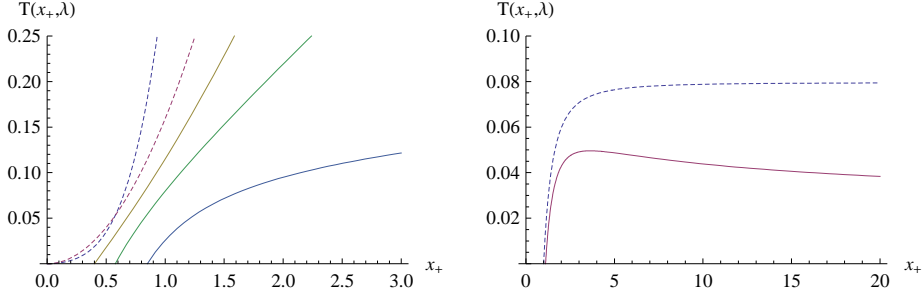}
\caption{Temperature  graphs of $T(x_\pm,\lambda)$ with $x_+=r_+$
and $l=1$. Left graphs for Lifshitz black holes with $1/3\le
\lambda <3$: $T(x_\pm,\lambda)$ for $\lambda=1/3,1/2,0.7,1,2$ from
top to bottom. We observe that for extremal points,
$T(0,1/3)=T(0,1/2)=T(0.40,0.7)=T(0.58,1)=T(0.85,2)=0$. Right graph
for Reissner-Nordstr\"om type black hole with $\lambda \ge 3$:
$T(x_\pm,\lambda)$ for $\lambda=3,4$. $T(1,3)=T(1.11,4)=0$ at
extremal points and  $T_m(3.63,4)=0.05$ at the maximum point.}
\label{fig.2}
\end{figure}

In order to explore the properties of Ho\v{r}ava-Lifshitz back
holes, let us first study the Hawking tempearture of these black
holes because it is independent of  the mass definition. The
temperature for $\lambda\not=1$ is defined  by
\be
\label{temp}T=\frac{(\tilde{N}^2f)'}{4\pi}\left.\sqrt{-g^{tt}g^{rr}}\right|_{x=x_+}=
\frac{x_+^{q(\lambda)}\Big[(2-p(\lambda))x_+^2 -p(\lambda)\Big]}
{4 \pi r_+}, \ee
while for $\lambda=1$ and SAdS black hole, they lead to \be
T_{\lambda=1}=\frac{3x_+^2-1} {8 \pi
r_+},~~T_{SAdS}=\frac{3x_+^2+2} {8 \pi r_+}. \ee Firstly,  we find
the extremal black holes from the condition of $T=0$ as \be
\label{extrecon} x_e=0,~{\rm for}~\frac{1}{3}\le \lambda \le
\frac{1}{2};~~
x_e=\sqrt{\frac{2\lambda-\sqrt{6\lambda-2}}{-2+\sqrt{6\lambda-2}}},~{\rm
for}~\lambda
> \frac{1}{2}. \ee As is shown in Fig. 2, $T$ is completely
different from $T_{SAdS}$  because the former is qualitatively
similar to the BTZ black hole with $T=0$ extremal
temperature~\cite{BTZ,BHTZ}, while the latter has a minimum
temperature $T_{\rm
min}(\sqrt{\frac{2}{3}})=\frac{\sqrt{3}}{2\sqrt{2}\pi}$
~\cite{CTZ,myung2} with its shape of $ \smile$.
 For $\lambda=1$, there
exists an extremal point  at $x_e=1/\sqrt{3}$  which the
temperature vanishes.  Hence, for $1/2 <\lambda < 3$, the solution
interpolates between AdS$_2\times S^2$ in the near-horizon
geometry of extremal black hole and Lifshitz at asymptotic
infinity.  On the other hand, for $1/3 \le \lambda \le 1/2$, one
could not have this connection because the corresponding black
hole  is a massless black hole located at the origin of coordinate
$x_e=0$  which is similar to the non-rotating BTZ black
hole~\cite{myung1}, even though they  are asymptotically Lifshitz.

Secondly, from the condition of $dT/dr_+=0$, we can easily
estimate  asymptotic behavior of the temperature because it
provides a checking point to test the presence of maximum
temperature. As is shown in Fig. 2,  we find the constant
asymptotic behavior of temperature at \be \lambda=3\ee and thus,
the maximum temperature always exists for \be \lambda >3.\ee  For
$\lambda>3$, we find completely different temperature because
their asymptotic behavior is similar to that of
Reissner-Norstr\"om black hole. We remind that $\lambda=3(z=0)$ is
the edge  of  Lifshitz black holes and thus,  it is not strange
for $\lambda>3$ to find the Reissner-Norstr\"om-type black hole in
asymptotically flat spacetimes.

It is not clear that the ADM mass for Ho\v{r}ava-Lifshitz black
holes  could be derived from mass parameter $m$ using the
Hamiltonian formulation without ambiguity.  The reason is because we
do not know clearly  what is asymptotically Lifshitz with dynamical
exponent $z \ge 0$~\footnote{However, recently, a boundary
stress-energy tensor approach  was used to derive the ADM mass of
Lifshitz black holes, especially for 3D Lifshitz black hole from the
new massive gravity~\cite{HT}.}. Up to now, we may calculate the ADM
mass only for asymptotically flat and anti-de Sitter spacetimes.
However, Cai-Cao-Ohta have tried to derive masses of LMP black holes
and its topological black holes~\cite{CCO2}. We also use this method
to derive the mass as
 \bea
M(r_+,\lambda)&=&\frac{\pi \kappa^2 \mu^2}{2\sqrt{6\lambda-2}}
(-\Lambda)^{2p(\lambda)+\frac{q(\lambda)}{2}} m^{2p(\lambda)} \nn \\
&=&\frac{\pi \kappa^2 \mu^2\sqrt{-\Lambda}}{2\sqrt{6\lambda-2}}
\frac{(1+x_+^2)^2}{x_+^{2p(\lambda)}}.\eea As is depicted in Fig.
3, we have mass  for Lifshitz black holes with $1/3\le \lambda
<3$.  For $1/3 \le \lambda \le 1/2$, there is no minimum point
which satisfies $dM/dx_+=0$ except at $x_+=0$, implying the
non-rotating BTZ black hole. For $\lambda=3,4$,  we find the mass
behavior  which is similar to that of  Ressner-Nordstr\"om  black
holes.

\begin{figure}[t!]
   \centering
   \includegraphics{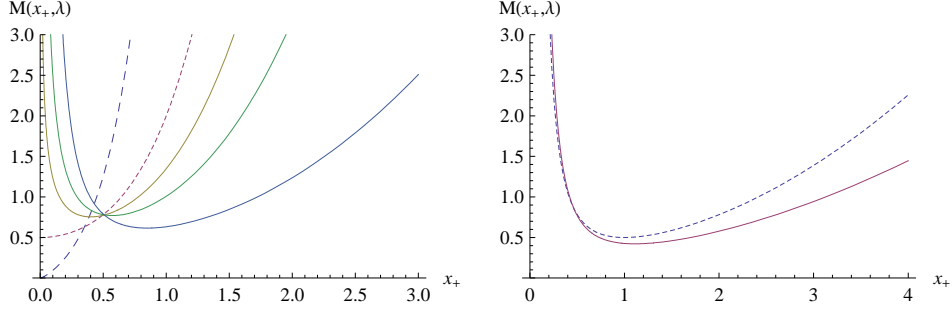}
\caption{Mass graphs of $M(x_\pm,\lambda)$. Left graphs for
Lifshitz black holes: $M(x_\pm,\lambda)$ for
$\lambda=2,1,0.7,1/2,1/3$ from top to bottom (along $M$-axis). For
$1/3 \le \lambda \le 1/2$, there is no minimum point which
satisfies $dM/dx_+=0$ except at $x_+=0$. Right graph for
Ressner-Nordstr\"om type black holes: $M(x_\pm,\lambda)$ for
$\lambda=3,4$.} \label{fig.3}
\end{figure}

\begin{figure}[t!]
   \centering
   \includegraphics{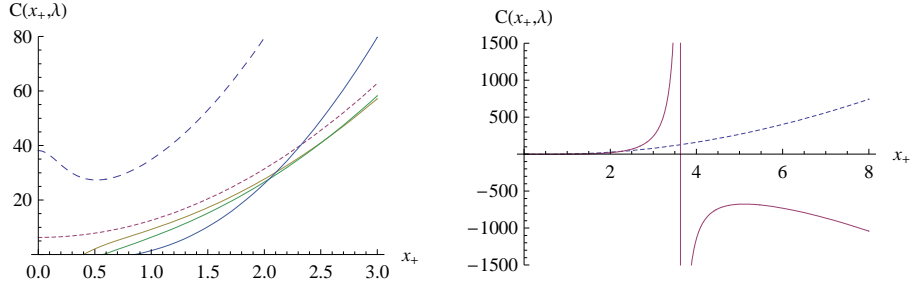}
\caption{Heat Capacity  graphs of $C(x_\pm,\lambda)$. Left graphs
for Lifshitz black holes: $C(x_\pm,\lambda)$ for
$\lambda=1/3,1/2,0.7,1,2$ from top to bottom. We check that for
extremal black holes, $C(0.4,0.7)=C(0.58,1)=C(0.85,2)=0$. Right
graph for Reissner-Norstr\"om type black holes: $C(x_+,4)$ shows a
blow-up point at $x_m=3.63$, which shows a feature of transition
from stable region to unstable region. $C(x_+,3)$ is an increasing
function of $x_+$. } \label{fig.4}
\end{figure}
The heat capacity is an important thermodynamic quantity which
tell us the stability of the LMP black holes. A heat capacity
defined by $C= \Big(\frac{dM}{dT}\Big)_\lambda$ takes the form
\begin{eqnarray}
C(x_+,\lambda)&=& \frac{4\pi^2\kappa^2\mu^2}{\sqrt{6\lambda-2}}
         \left(  \frac{(1+x^2_+)[(2-p(\lambda))x^2_+-p(\lambda)]}
           {(2-p(\lambda))(1+q(\lambda))x^2_++p(\lambda)(1-q(\lambda))}\right),
\end{eqnarray}
while for $\lambda=1$ and SAdS cases, they have
 \be C_{\lambda=1}(x_+)=2 \pi^2\kappa^2\mu^2(1+x^2_+)
\Bigg(\frac{3x^2_+-1}{3x^2_++1}\Bigg),~~C_{SAdS}=8\pi x^2_+
\Bigg(\frac{3x^2_++2}{3x^2_+-2}\Bigg). \ee As is shown in Fig. 4,
for Lifshitz black holes,  all heat capacities  go to $\infty$ as
$x_+ \to \infty$.  We observe that the Ho\v{r}ara-Lifshitz (LMP)
black hole with $1/3 \le \lambda < 3$ are always thermodynamically
stable because their heat capacities are always positive for
$x_+\ge x_e$. On the other hand, for $\lambda >3$, the heat
capacity blows up at the minimum point of $x_m=3.63$. Hence, small
black holes of $x_+<x_m$ are stable because  their heat capacities
are positive, while large black hole of $x_+>x_m$ are unstable
because their heat capacities are negative. This indicates a
feature of Reissner-Norstr\"om type black holes.

Finally, the entropy of  Ho\v{r}ava-Lifshitz black holes could be
found by assuming that  the first law of thermodynamics holds \be
dM=TdS. \ee Using this first law, we derive an entropy
\begin{equation}
S=\frac{\sqrt{2}(-\Lambda)}{\sqrt{3\lambda-1}}\left(\frac{A}{4}-\frac{\pi}{\Lambda}\ln\frac{A}{4}-S_0\right),
\end{equation}
with $A/4=\pi r_+^2$ and  $S_0=\frac{\pi\ln\pi}{-\Lambda}$.
\begin{figure}[t!]
   \centering
   \includegraphics{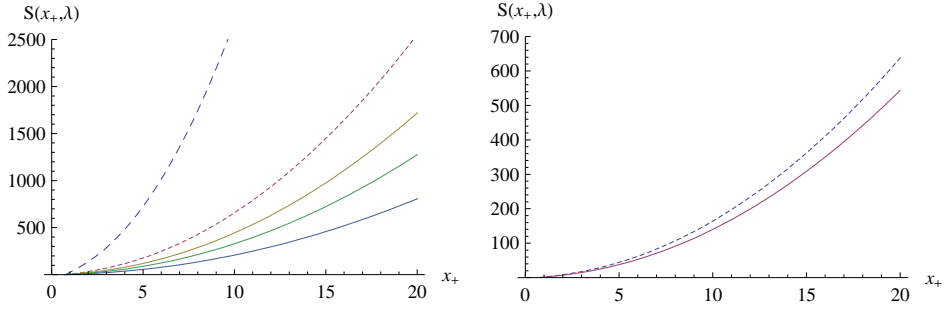}
\caption{Entropy graphs of $S(x_\pm,\lambda)$. Left graphs for
Lifshitz black holes: $S(x_\pm,\lambda)$ for
$\lambda=1/3,1/2,0.7,1,2$ from top to bottom.  Right graph for
Reissner-Norstr\"om type black holes: for $\lambda=3,4$ from top
to bottom,
 $S(x_+,\lambda)$ shows similar behaviors as the Lifshitz black holes show.} \label{fig.5}
\end{figure}
The main feature of the LMP black holes has the entropy with a
logarithmic term~\cite{CCO,CCO2}. As is shown in Fig. 5, there is
no drastic change in the  entropy when changing from Lifshitz to
Reissner-Norstr\"om type black holes.

However, for $ \lambda >3$, we have a different form of the
entropy
\begin{equation}
S=\frac{\sqrt{2}(-\Lambda)}{\sqrt{3\lambda-1}}\left(\frac{A}{4}+\pi
Q^2\ln\frac{A}{4}-S_0\right)
\end{equation}
because we may interpret $\frac{1}{-\Lambda_w}$ to be the charge
of $Q^2$ in the Reissner-Norstr\"om type black holes.

\section{Discussions}

We study thermodynamics of the LMP black holes in the $z=3$
Ho\v{r}ava-Lifshitz gravity  according to the coupling constant
$\lambda$.
  For $1/3 \le \lambda \le 1/2$, the black holes behave the
  non-rotating BTZ back hole because of their extremal point at
  $x_e=0$, while for $ 1/2 < \lambda < 3$ including $\lambda=1$, the black holes behave the
  rotating BTZ back hole because of their extremal point at
  $x_e\not=0$. However,  all these black holes belong to asymptotically Lifshitz.
For $\lambda >3$, we have found Reissner-Norstr\"om type black
holes. We confirm this classification by studying thermodynamic
properties of LMP black holes.

Even though we have started with the $z=3$ Ho\v{r}ava-Lifshitz
gravity, we have Lifshitz black hole with dynamical exponent $z$ and
Reissner-Norstr\"om type black holes. Frankly speaking, as far as is
concerned on obtaining the LMP black holes,  we were working with
$z=2$  Ho\v{r}ava-Lifshitz gravity without the Cotton tensor
($C_{ij}=0$).  We suggest that  various black hole solutions are
found  from this gravity because it  is a non-relativistic gravity
theory.

On the other hand, there are Lifhsitz black holes obtained from
relativistic higher derivative gravities, where the higher
derivative terms are not considered as  perturbative corrections to
Einstein-Hilbert action. These are included $z=3$ Lifshitz black
hole from the 3D new massive gravity~\cite{AGGH}, $z=3/2$ Lifshitz
black hole in 4D spacetimes~\cite{CLS}, and higher dimensional
Lifshitz black holes~\cite{AGGHa}.

Consequently, our main results are  the interpretation of  the
Ho\v{r}ava-Lifshitz (LMP) black holes with $\lambda \ge 1/3$ as
the Lifshitz black holes with dynamical exponent $z$ for $ 1/3 \le
\lambda < 3(0< z \le 4)$ and the Reissner-Nordstr\"om black hole
with charge $\frac{1}{-\Lambda_W} $ for $\lambda
>3$.

\section*{Acknowledgement}
The authors thank Hyung Won Lee for helpful discussions.  Y. S.
Myung was supported by the National Research Foundation grant funded
by the Korea government(MEST) through the Center for Quantum
Spacetime (CQUeST) of Sogang University with grant number
2005-0049409.  Y.-W. Kim was supported by the Korea Research
Foundation Grant funded by Korea Government (MOEHRD):
KRF-2007-359-C00007.

\end{document}